\documentclass[seceq]{ptptex}

\usepackage{graphicx}

\title{
Formation of $\phi$ mesic nuclei}%

\author{
Junko \textsc{Yamagata-Sekihara}$^1$\footnote{Departamento de F\'{\i}sica Te\'orica and IFIC,
 Universidad de Valencia-CSIC, E-46071 Valencia, Spain}, Daniel \textsc{Cabrera}$^2$, Manuel J. \textsc{Vicente Vacas}$^3$
and Satoru \textsc{Hirenzaki}$^4$%
}

\inst{
$^1$Yukawa Institute for Theoretical Physics, Kyoto University, Kyoto 606-8502, Japan\\
$^2$Departamento de F\'{\i}sica Te\'orica II,
Universidad Complutense, 28040 Madrid, Spain\\
$^3$Departamento de F\'{\i}sica Te\'orica and IFIC,
 Universidad de Valencia-CSIC, E-46071 Valencia, Spain\\
$^4$Department of Physics, Nara Women's University, Nara 630-8506, Japan \\
}

\abst{
We study the structure and formation of the $\phi$ mesic nuclei to investigate the in-medium modification of the $\phi$-meson spectral function at finite density.
We consider (${\bar p},\phi$), ($\gamma,p$) and ($\pi^-,n$) reactions to produce a $\phi$-meson inside the nucleus and evaluate the effects of its medium modifications to the reaction cross sections.
We also estimate the consequences of the uncertainties of the ${\bar K}$ selfenergy in medium to the $\phi$-nucleus interaction.
We find that it may be possible to see a peak structure in the reaction spectra for the strong attractive potential cases.
On the other hand, for strong absorptive interaction cases with relatively weak attractions, it is very difficult to observe clear peaks and we may need to know the spectrum shape in a wide energy region to deduce the properties of $\phi$.
}

\begin{document}

\maketitle


\section{Introduction}
\label{sec:1}
Mesic atoms such as pionic- and kaonic-atoms are Coulomb assisted meson-nucleus bound systems and have been studied systematically for a long time~\cite{ref:1}. The binding energies and widths of these bound states provide us unique and valuable information on the meson-nucleus interactions.  Mesic nuclei, another kind of meson-nucleus systems, are meson-nucleus bound states mainly due to attractive strong interactions and have also been studied intensively in these days~\cite{gata06,YamagataSekihara:2008ji,Ikuta:2002kr,Koike:2007iz,nagahiro05,jido02}.
In the contemporary hadron-nuclear physics, mesic atoms and mesic nuclei are considered to be very interesting objects in the following two aspects. First, they are {\it strongly interacting exotic many body systems} which should be explored by nuclear physicists. 
Second, mesic atoms and mesic nuclei provide unique laboratories for the studies of {\it hadron properties at finite density} which are significantly important to explore various aspects of the symmetries of the strong interaction~\cite{ref:2,ref:3}. 

In this letter, we report our studies on structure and formation of $\phi$ mesic nuclei.  
From the study of the $\phi$ meson properties in nucleus, we can intuitively expect to obtain new information on the OZI rule \cite{ozi} and the ${\bar s}s$ component of the nucleon at finite density.
The analysis of QCD sum rules~\cite{ref:5} and the data taken at KEK~\cite{ref:6} suggested a 3
$\%$ mass reduction of $\phi$ at normal nuclear density.
On the other hand, in-medium properties of the $\phi$ meson have also been studied theoretically with a close relation to $K$ and $\bar{K}$ meson properties in medium because of the strong $\phi \rightarrow  K\bar{K}$ coupling.
The $\phi$ meson selfenergy calculated in Refs.~\citen{ref:7,ref:8} indicate a significantly smaller attractive potential for $\phi$. 
Since we can expect to obtain new information which is complementary to the invariant mass measurements~\cite{ref:6}, we investigate the $\phi$ mesic nuclear states to study the $\phi$ properties in medium. 
Especially, in the experiment~\cite{ref:6}, only a small fraction of the produced $\phi$ meson decays inside the nucleus because of the long life time of the $\phi$ meson.
In the $\phi$ mesic nucleus, all $\phi$ mesons must stay inside the nucleus until their decay or absorption.
Thus, we may have possibilities to get better information on $\phi$ properties in nucleus.

We show the calculated results of the bound states and formation spectra for some reactions in the following sections which will help to consider experimental feasibilities.

\section{Optical potential and bound states}
\label{sec:2}
We study the properties of the $\phi$ mesic nuclei by solving the Klein-Gordon equation,
\begin{equation}
[-\vec{\nabla}^2+\mu^2+\Pi_\phi(E,\rho(r))]\phi({\vec r})=\omega^2\phi({\vec r})~~.
\label{eq:1}
\end{equation}
Here, $\mu$ is the $\phi$-nucleus reduced mass.
We employ the empirical Woods-Saxon form for the nuclear density as,
\begin{equation}
\rho(r)=\rho_p(r)+\rho_n(r)=\displaystyle{\frac{\rho_0}{1+\exp[(r-R)/a]}}
\label{eq:2}
\end{equation}
where $R$ and $a$ indicate radius and diffuseness parameters for the nucleus, and are fixed to be $R=1.18A^{1/3}-0.48$~(fm) and $a=0.5$~(fm), respectively.
We solve the Klein-Gordon equation numerically following the method of Oset and Salcedo~\cite{Oset:1985tb}.

\begin{figure}
  \includegraphics[width=1\textwidth]{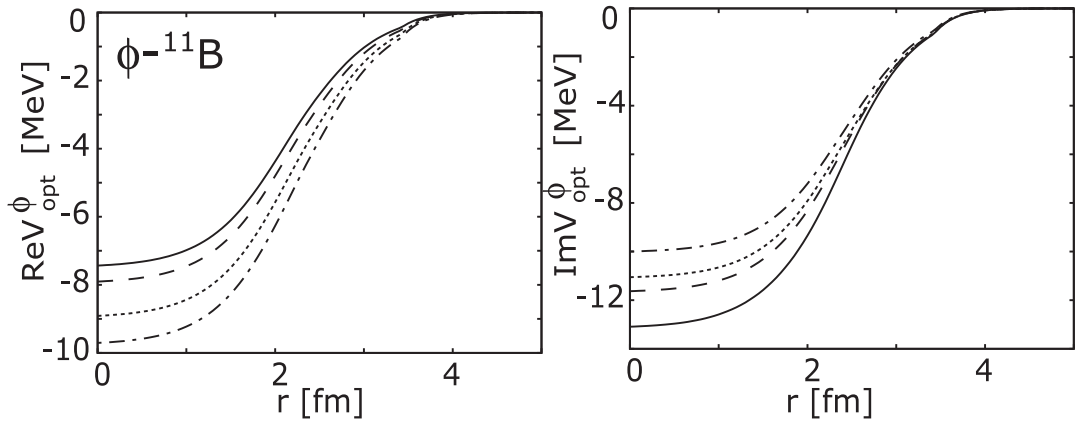}
\caption{ The $\phi$ meson optical potentials as a function of the radial coordinate $r$ for $\phi$ -$^{11}$B systems obtained from $\Pi_\phi$ in Ref.~\citen{ref:8}.
The left and right panels show the real and imaginary parts.
The solid, dashed, dotted, dotted-dashed lines indicate the potential strength for the $\phi$ meson energies $E-m_\phi=0$ MeV, $-10$ MeV, $-20$ MeV and $-30$ MeV, respectively.}
\label{fig:Vopt}       
\end{figure}

We show the $\phi$ meson optical potential $V_{\rm opt}^\phi(r,E)$ in Fig.~\ref{fig:Vopt}, which is obtained from the $\phi$ meson selfenergy $\Pi_\phi$ in Ref.~\citen{ref:8} as $V_{\rm opt}^\phi(r,E)=\displaystyle{\frac{1}{2\mu}}\Pi_\phi(E,\rho(r))$ within the local density approximation.
The selfenergy $\Pi_\phi$ depends on the energy $E$ of the $\phi$ meson and on the density $\rho$.

As shown in Fig.~\ref{fig:Vopt}, the theoretical potential based on $\Pi_\phi$~\cite{ref:8} is a weak attractive potential, Re $V_{\rm opt}^\phi(0,E=m_\phi)\sim -7.5$ MeV.
The absorptive part of the potential ${\rm Im}~V_{\rm opt}^\phi(r,E)$ is also weak and has a similar strength as the real part.
The theoretical model of Ref.~\citen{ref:8} is based on the OZI rule~\cite{ozi}, where the nuclear medium effects of the $\phi$ meson are mainly from those of $K$ and ${\bar K}$ through the strong $\phi K{\bar K}$ coupling.
Hence, it seems natural that the $\phi$-nucleus potential is weak.
On the other hand, a 3\% mass reduction is indicated in Refs.~\citen{ref:5,ref:6}.
To simulate it, we multiply by a factor the real part of the theoretical potential so that Re $V_{\rm opt}^\phi(0,E=m_\phi)$ scales to $-30$ MeV.
We use both theoretical (shallow) and scaled (deep) potentials and compare the results to know the sensitivity of the observables to the potential depth.

We calculate the binding energies B.E. and widths $\Gamma$ of the bound states, which are related to the eigen energy $\omega$ in Eq.~(\ref{eq:1}) as $\omega=\mu-{\rm B.E.}-i\Gamma/2$, by solving the Klein-Gordon equation selfconsistently for the real part of the $\phi$ meson energy to know the resonance energies in the reaction spectra observed in experiments.
The calculated results are shown in Table~\ref{tab:1} for four nuclei which are expected to be in the final states in the one proton pick-up reactions considered in the next section.
\begin{table} 
\begin{center}
\caption{Calculated binding energies and widths of $\phi$ meson bound states in $^{11}$B,~$^{39}$K,~$^{123}$In,~and~$^{207}$Tl.
Widths do not include the $\phi$ decay width in vacuum $\Gamma_\phi^{\rm free}=4.26$~MeV~\cite{pdg} in this Table.
}
\label{tab:1}       
\begin{tabular}{lll}
\hline\noalign{\smallskip}
 & Shallow potential & Deep potential \\
 & B.E.[MeV]($\Gamma$[MeV]) & B.E.[MeV]($\Gamma$[MeV]) \\
\noalign{\smallskip}\hline\noalign{\smallskip}
$^{11}$B & none & $1s$~~10.2(17.9) \\
$^{39}$K & none & $1s$~~28.5(20.0) \\
& & $2p$~~11.8(19.8) \\
$^{123}$In & $1s$~~2.34(21.6) & $1s$~~34.5(18.7) \\
& & $2s$~~12.2(19.9) \\
& & $2p$~~26.3(19.9) \\
$^{207}$Tl&$1s$~~3.73(22.5) & $1s$~~40.1(17.5) \\
& & $2s$~~24.4(20.2) \\
& & $2p$~~27.1(20.3) \\
& & $3p$~~13.1(20.4) \\
& & $3d$~~27.5(19.8) \\
\noalign{\smallskip}\hline
\end{tabular}
\end{center}
\end{table}
We see that there exist several bound states of $\phi$ in nuclei, however, the widths of the states are large and will make it difficult to observe any peak structure in the missing mass spectra.

\section{Formation spectra of $\phi$ mesic nuclei}
\label{sec:3}
We consider three types of proton pick-up reactions to form the $\phi$-mesic nucleus which are ($\gamma,p$), ($\pi^-,n$) and (${\bar p},\phi$).
The elementary processes of these three reactions are $\gamma+p\to\phi+p$~\cite{mibe}, $\pi^-+p\to \phi+n$~\cite{Dahl:1967pg}, and ${\bar p}+p\to \phi+\phi$~\cite{Evangelista:1998zg}, respectively.
We would like to make a few comments on the data in Ref.~\citen{mibe}, which have not been published yet.
In the present calculation, we need to use the backward cross section of $\phi$ production in the elementary process, which was evaluated here by a functional fit to the data shown in Ref.~\citen{mibe}.
However there are large uncertainties in the value of the backward cross section that could affect the absolute value but not the shape of the spectrum.

\begin{figure}[htbp]
\begin{center}
  \includegraphics[width=0.5\textwidth]{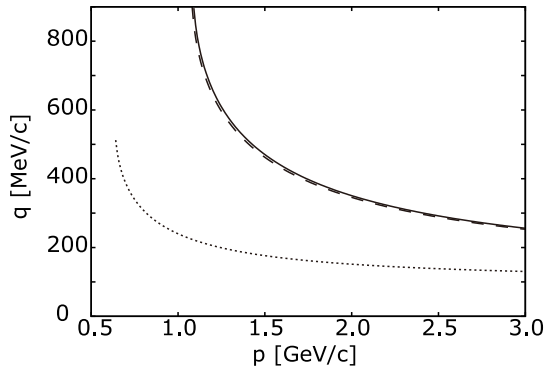}
\caption{Momentum transfer of the ($\gamma,p$) (solid curve), ($\pi^-,n$) (dashed curve) and (${\bar p},\phi$) (dotted curve) reactions for the $\phi$ mesic nucleus formation.
Protons in the initial state are assumed to be in $1p_{3/2}$ state in $^{12}$C.}
\label{fig:mom}       
\end{center}
\end{figure}

We first show in Fig.~\ref{fig:mom} the momentum transfer of the reactions as it is an important information of meson bound state formation~\cite{Toki:1990fc,Hirenzaki:1991us}.
We found that the momentum transfer of the (${\bar p},\phi$) reaction is relatively smaller than the other two reactions.
However, we also found that the momentum transfer is always larger than 100 MeV/c in the energy range plotted in Fig.~\ref{fig:mom}.

\begin{figure}[htbp]
\begin{center}
  \includegraphics[width=0.7\textwidth]{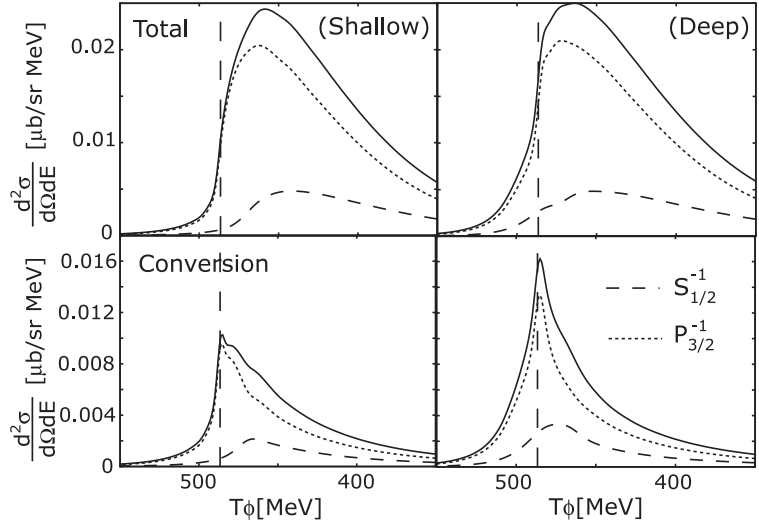}
\caption{Calculated spectra for the $\phi$-nucleus systems formation plotted as a function of the emitted $\phi$ meson energy in the $^{12}$C(${\bar p},\phi$) reaction at $p_{\bar p}=1.3$ GeV/c.
Total cross sections (upper panels) and conversion parts (lower panels) are calculated with the theoretical (shallow) optical potential (left panels) and the scaled (deep) optical potential (right panels).
The dashed, dotted lines show the contributions from different proton hole states as indicated in the figure.
The vertical dashed line indicates the $\phi$ meson production threshold.}
\label{fig:spec}       
\end{center}
\end{figure}

We use the Green's function method to calculate the formation spectra of the $\phi$ meson from nucleus~\cite{morimatsu85} as in the cases of other mesic nucleus formation~\cite{gata06,YamagataSekihara:2008ji,Ikuta:2002kr,Koike:2007iz,nagahiro05,jido02}.
We briefly explain the formalism for the (${\bar p},\phi$) reaction as an example.
The expected spectra of the (${\bar p},\phi$) reactions $\displaystyle{\Bigl(\frac{d^2\sigma}{d\Omega dE_\phi}\Bigr)}$ are evaluated from the nuclear response function $S(E)$ and the elementary cross section $\displaystyle{\Bigl(\frac{d\sigma}{d\Omega}\Bigr)^{\rm ele}}$ in the impulse approximation as
\begin{equation}
\label{eq:6}
\displaystyle{\Bigl(\frac{d^2\sigma}{d\Omega dE_\phi}\Bigr)=\Bigl(\frac{d\sigma}{d\Omega}\Bigr)^{\rm ele}\times S(E)}~~.
\end{equation}

The calculation of the nuclear response function with a complex potential is formulated by Morimatsu and Yazaki~\cite{morimatsu85} as 
\begin{equation}
\label{S(E)}
S(E)=-\frac{1}{\pi} {\rm Im}\sum_f \int d{\vec r}d{\vec r}\,' \tau^\dagger_f G(E; {\vec r},{\vec r}\,') \tau_f,
\end{equation}
\noindent
where the summation is taken over all possible final states.
$G(E; {\vec r},{\vec r}\,')$ is the Green's function of $\phi$ meson interacting in the nucleus and defined as,
\begin{equation}
\label{Gfunc}
G(E; {\vec r},{\vec r}\,')=\langle\alpha |\phi({\vec r})\displaystyle{\frac{1}{E-H_\phi+i\epsilon}}\phi^+({\vec r}\,')|\alpha\rangle~~,
\end{equation}
where $\alpha$ indicates the proton hole state and $H_\phi$ indicates the Hamiltonian of the $\phi$ meson-nucleus system.
The amplitude $\tau_f$ denotes the transition of the incident particle (${\bar p}$) to the nucleon-hole and the outgoing $\phi$ meson, involving the nucleon-hole wavefunction $\psi_{j_N}$ and the distorted waves $\chi_i$ and $\chi_f$, of the projectile and ejectile.
By taking the appropriate spin sum, the amplitude $\tau_f$ can be written as,
\begin{equation}
\label{tau}
\tau_f({\vec r})=\chi_f^*({\vec r})\xi_{1/2,m_s}^*[Y_{l_\phi}^*(\hat{\vec r})\otimes \psi_{j_N}({\vec r})]_{JM}\chi_i({\vec r})~,~
\end{equation}
\noindent
with the meson angular wavefunction $Y_{l_\phi}(\hat{\vec r})$ and the spin wavefunction $\xi_{1/2,m_s}$ of the ejectile.
We assume the harmonic oscillator wavefunctions for $\psi_{j_N}$ with the empirical range parameter.

The
{semi-}exclusive spectra can be calculated by decomposing the response function~(\ref{S(E)}) into the escape and conversion parts: $S=S_{\rm esc}+S_{\rm con}$.
This decomposition can be done exactly by
\begin{eqnarray}
&&S_{\rm esc}(E)\nonumber\\
&&=-\displaystyle{\frac{1}{\pi}}\sum_f \int d{\vec r}d{\vec r}\,' \tau_f^\dagger (1+G^\dagger V_{\rm opt}^{\phi\dagger} (r,E)){\rm Im}G_0(1+V_{\rm opt}^\phi(r,E)G)\tau_f\nonumber\\
&&S_{\rm con}(E)=-\displaystyle{\frac{1}{\pi}}\sum_f  \int d{\vec r}d{\vec r}\,'\tau_f^\dagger G^\dagger {\rm Im}V_{\rm opt}^\phi(r,E)G\tau_f~~.
\label{eq:9}
\end{eqnarray}
\noindent
where $V_{\rm opt}^\phi(r,E)$ is the $\phi$ meson-nucleus optical potential given in the Hamiltonian.
The conversion part is known to express the contributions of the $\phi$ meson absorption to the (${\bar p},\phi$) spectra~\cite{morimatsu85}.

We show the calculated cross sections of the $^{12}$C(${\bar p},\phi$) reaction proposed in Ref.~\citen{iwasaki_pro} for the formation of a $\phi$ -$^{11}$B system in Fig.~\ref{fig:spec} for the theoretical (shallow) optical potential case in the left panels and for the scaled (deep) optical potential case in the right panels.
The conversion part of the spectra are also shown in the lower panels in addition to the total spectra in the upper panels.
The spectra shown in this article are folded spectra with a Gau$\beta$ian distribution with $\phi$ meson decay width in vacuum at $E=m_\phi$.
For the shallow optical potential case, since no bound state exists as shown in Table~\ref{tab:1}, the spectrum has a smooth shape without any peak structures in the bound region.
For the deep potential case, we find some enhancement of the spectra in the bound energy region.
However, we do not observe any peak structures again for the deeper potential case even when there exists one bound state.
We also find that the conversion contributions shown in lower panels, which correspond to the (${\bar p},\phi$) spectra coincident with the particle emissions from $\phi$ meson decay (absorption) in nucleus, resemble each other for both the shallow and deep potential cases, even though they have different constitutions of subcomponents.
The conversion part of the spectra was defined in Eq.~(\ref{eq:9}) as a contribution of the absorptive interaction ${\rm Im}~V_{\rm opt}^\phi(r,E)$.
One of the advantages of the (${\bar p},\phi$) reaction is a possible background reduction by the conversion spectrum as described in Ref.~\citen{iwasaki_pro}.
Thus, the conversion spectrum in Fig.~\ref{fig:spec} will be an important piece of information to observe the $\phi$ meson properties in nucleus.

\begin{figure}[htbp]
\begin{center}
  \includegraphics[width=1.0\textwidth]{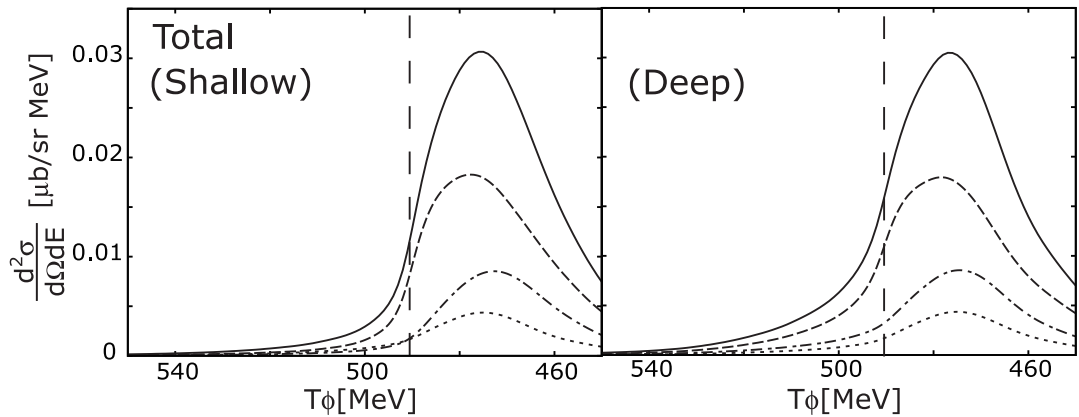}
\caption{Calculated spectra for the $\phi$-nucleus systems formation plotted as a function of the emitted $\phi$ meson energy in the $^{124}$Sn(${\bar p},\phi$) reaction at $p_{\bar p}=1.3$ GeV/c.
We include the contributions from outer proton orbits, $1g_{9/2}$ (dashed lines), $2p_{1/2}$(dotted lines), and $2p_{3/2}$(dash-dotted lines).
Total cross sections are calculated with the theoretical (shallow) optical potential (left panel) and the scaled (deep) optical potential (right panel).
The vertical dashed line indiates the $\phi$ meson production threshold.}
\label{fig:124Sn}       
\end{center}
\end{figure}

To investigated the mass number dependence of the formation spectra, we show the calculated spectra in Fig.~\ref{fig:124Sn} for $^{124}$Sn target with the shallow and deep potential cases.
We find that there appear some enhancements in the $\phi$ meson bound energy region for the deep potential case.
However, again we do not see any clear peaks corresponding to the bound states.
As we have expected from the widths of the bound states, it is difficult to observe any peak structure in the standard missing mass spectra.

\begin{figure}[htbp]
\begin{center}
  \includegraphics[width=0.7\textwidth]{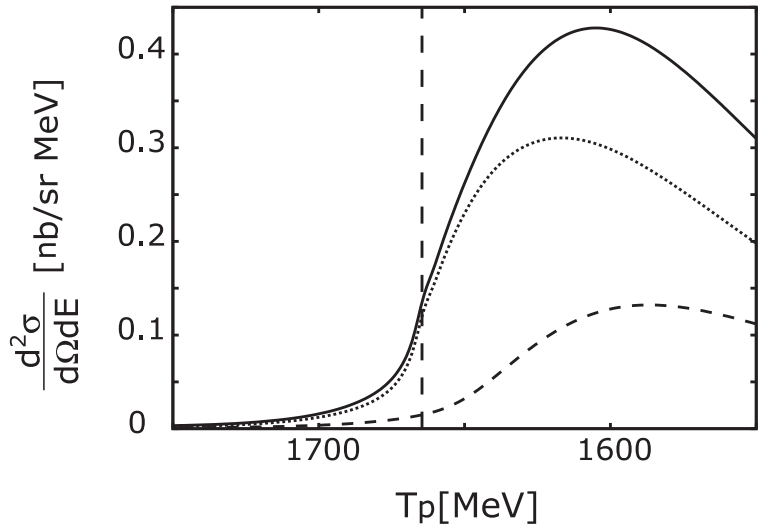}
\caption{Calculated spectra for the $\phi$-nucleus systems formation plotted as a function of the emitted proton energy in the $^{12}$C($\gamma,p$) reaction at $p_{\gamma}=2.7$ GeV/c.
The dashed, dotted lines show the contributions from $s_{1/2}^{-1}$ and $p_{3/2}^{-1}$ proton hole states, respectively.
The vertical dashed line indicates the $\phi$ meson production threshold.
The theoretical (shallow) optical potential is used.
The absolute value of this spectrum is still ambiguous. See detail in text. }
\label{fig:gamma_chiral}       
\end{center}
\end{figure}

\begin{figure}[htpb]
\begin{center}
  \includegraphics[width=0.7\textwidth]{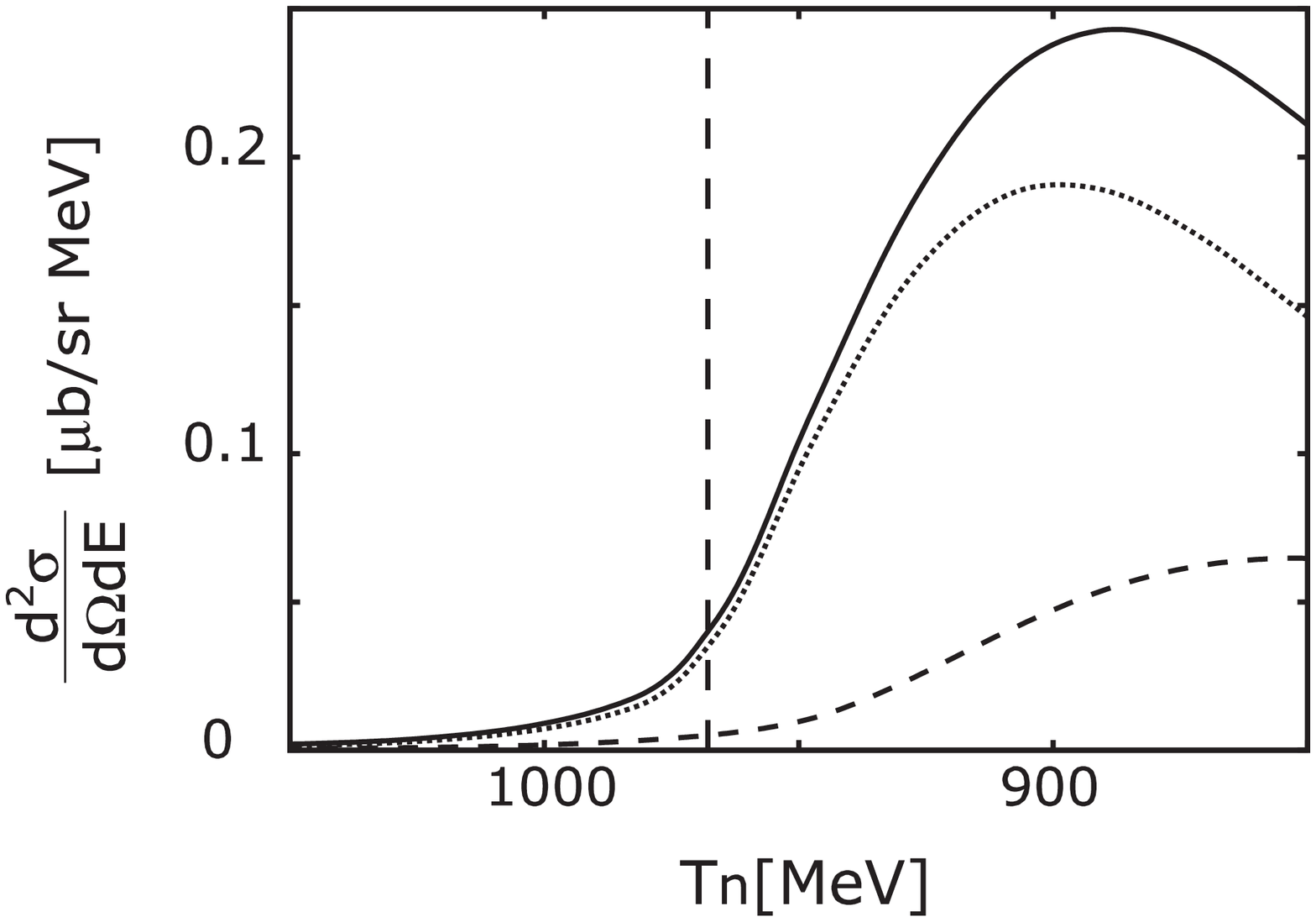}
\caption{Calculated spectra for the $\phi$-nucleus systems formation plotted as a function of the emitted neutron energy in the $^{12}$C($\pi^-,n$) reaction at $p_{\pi^-}=2.0$ GeV/c.
The dashed, dotted lines show the contributions from $s_{1/2}^{-1}$ and $p_{3/2}^{-1}$ proton hole states, respectively.
The vertical dashed line indicates the $\phi$ meson production threshold. 
The theoretical (shallow) optical potential is used.}
\label{fig:pin_chiral}       
\end{center}
\end{figure}

In Figs.~\ref{fig:gamma_chiral} and \ref{fig:pin_chiral}, we show the calculated results for different formation reactions, namely, $^{12}$C($\gamma,p$)~(Fig.~\ref{fig:gamma_chiral}) and $^{12}$C($\pi^-,n$) (Fig.~\ref{fig:pin_chiral}).
In both cases, the momentum transfers of the reactions are larger than for the ${\bar p}$ induced case and, hence, the spectra around threshold are suppressed in these reactions.

\begin{figure}[htpb]
\begin{center}
  \includegraphics[width=1.0\textwidth]{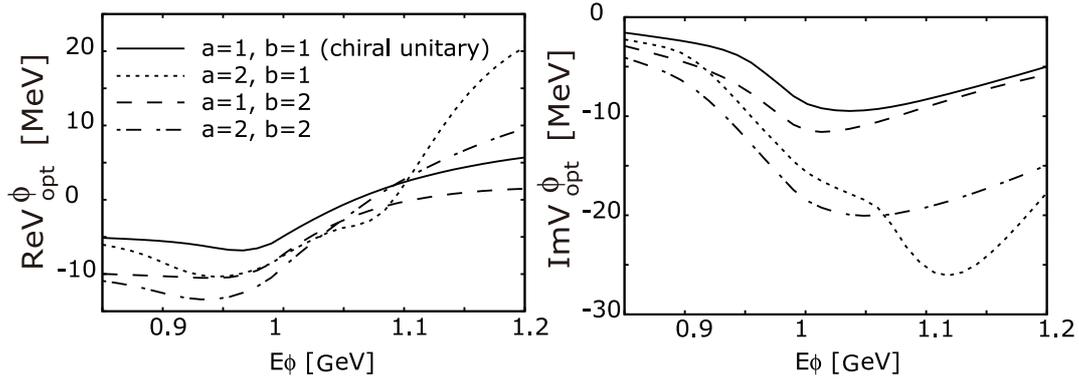}
\caption{Real part (left) and imaginary part (right) of the calculated $\phi$-nucleus optical potential $V_{\rm opt}^\phi(r,E)=\Pi_\phi(E_\phi,\rho(r))/2\mu$ are plotted as a function of the $\phi$ meson energy $E_\phi$ at $\rho(r)=\rho_0$.
Each line shows the $V_\phi$ obtained with different $\Pi_{K^-}$ in the model in Ref.~\citen{ref:8}.
Please see detail in text.}
\label{fig:pi}       
\end{center}
\end{figure}

We have checked the sensitivity of the optical potential $V_{\rm opt}^\phi(r,E)$ to the $K^-$ selfenergy $\Pi_{K^-}$ in the nucleus.
Since in the theoretical model of Ref.~\citen{ref:8}, the medium modification of the $K^+$ and $K^-$ cloud around the $\phi$ meson dominantly  causes the modifications of the $\phi$ properties, and since the $K^-$ selfenergy $\Pi_{K^-}$ has a large uncertainty, it is interesting to change $\Pi_{K^-}$ and study the effects in $\Pi_\phi$.
We simply parametrize the modified $K^-$ selfenergy $\tilde{\Pi}_{K^-}$ as 
\begin{equation}
\tilde{\Pi}_{K^-}=a({\rm Re}~\Pi_{K^-})+b({\rm Im}~\Pi_{K^-})~~,
\label{eq:pi}
\end{equation}
where the $\Pi_{K^-}$ in the right hand side is obtained by the chiral unitary model in Ref.~\citen{ref:8}.
We consider the constant parameters $a$ and $b$ to scale the real and imaginary parts of $\Pi_{K^-}$.
We show the calculated $\Pi_{\phi}$ for four cases with $a,~b=1$ and/or $2$ in Fig.~\ref{fig:pi}.
We find that a certain ambiguity of $\Pi_\phi$ due to that of $\Pi_{K^-}$ exists as we expected.
However, the strength of the real part and imaginary part of $\Pi_\phi$ are correlated, namely, the strong attractive real part always appears with the strong absorptive imaginary part and so on.
Thus, in any case, it will be difficult to see any clear structure well separated from the threshold and the quasi-elastic contribution in the formation cross section.

\begin{figure}[htpb]
\begin{center}
  \includegraphics[width=1.0\textwidth]{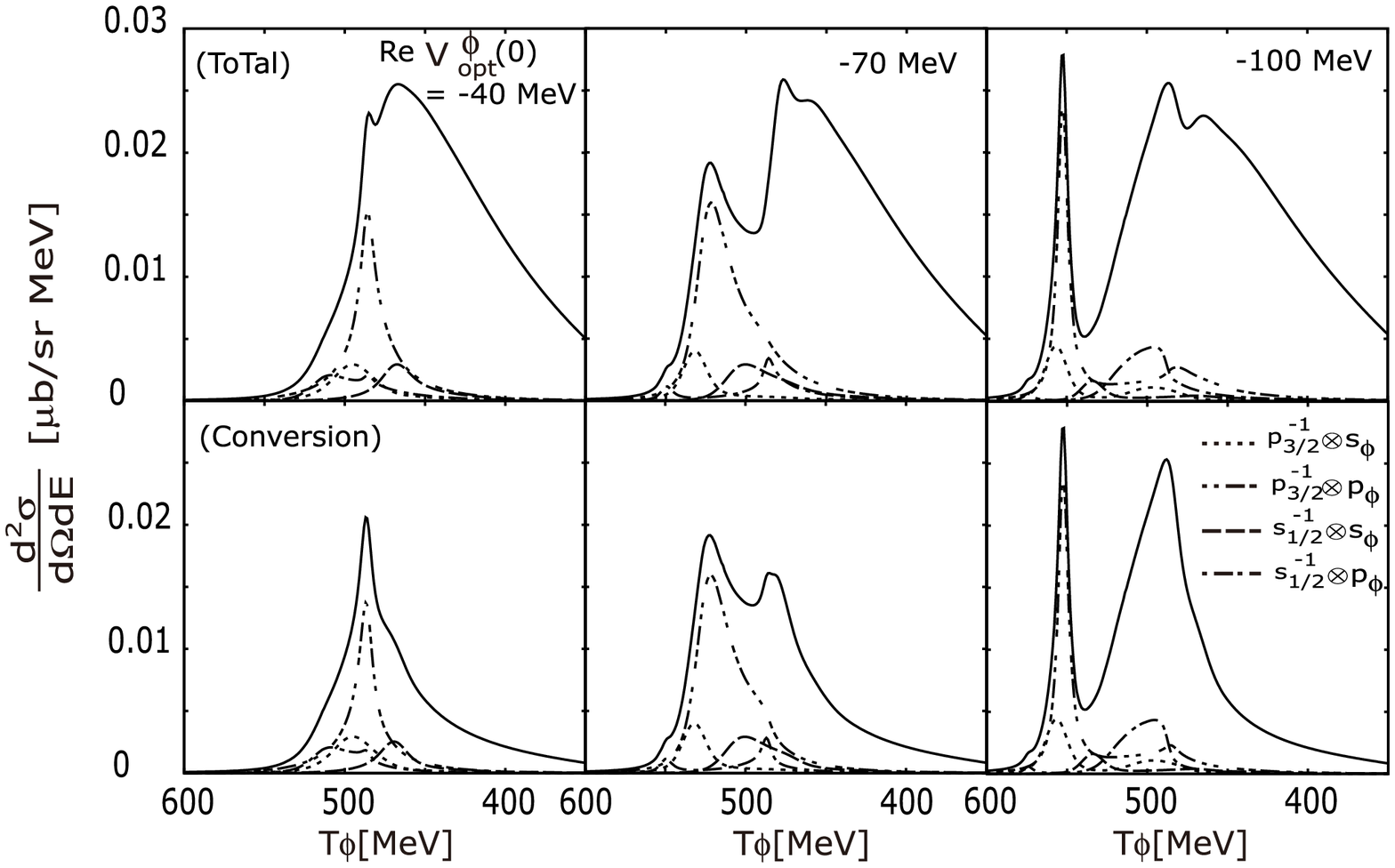}
\caption{Calculated spectra for the $\phi$-nucleus systems formation plotted as function of the emitted $\phi$-meson energy in the $^{12}$C(${\bar p},\phi$) reaction at $p_{\bar p}=1.3$ GeV/c.
Total cross section (upper panels) and conversion parts (lower panels) are calculated with deep potentials obtained by scaling the theoretical potential~\cite{ref:8} as ${\rm Re}~V_{\rm opt}^\phi(0,E=m_\phi)=-40,~-70$, and $-100$ MeV while keeping the imaginary part unmodified.
The dashed and dotted lines indicate the contributions from different sub components [$l^{-1}_j\otimes l_\phi$] as shown in the figure.}
\label{fig:7}       
\end{center}
\end{figure}

Finally, we consider deep potentials to check the possibility to observe a peak structure.
We scale again the theoretical potential in Ref.~\citen{ref:8} and consider deep real potential cases as ${\rm Re}~V_{\rm opt}^\phi(0,E=m_\phi)=-40,~-70$, $-100$ MeV keeping the ${\rm Im}~V_{\rm opt}^\phi(r,E)$ to the theoretical value.
The calculated results are shown in Fig.~\ref{fig:7}.
We find that we can observe a clear peak structure if the optical potential is as strongly attractive as $|{\rm Re}~V_{\rm opt}^\phi(0,E=m_\phi)|\geq 70$ MeV.

\section{Summary}
In this article we have studied the structure and formation of the $\phi$ mesic nuclei to investigate the in-medium modifications of the $\phi$ meson properties.
We have calculated the binding energies and the absorption widths of the $\phi$ meson bound states.
We have also studied the production of the $\phi$ meson in the nucleus and obtained numerical results for the formation spectra of the (${\bar p},\phi$), ($\gamma,p$) and ($\pi^-,n$) reactions.
Due to the large decay width of the $\phi$ meson in nucleus, we found that the observation of a clear peak in the missing mass spectra will be difficult if the $\phi$-nucleus potential is weakly attractive as reported in Refs.~\citen{ref:7,ref:8}.
A deep potential (${\rm Re}~V_{\rm opt}^\phi(0,E=m_\phi)\sim -30$~MeV) corresponding to the $3~\%$ mass reduction reported in Refs. \citen{ref:5,ref:6} is not enough to produce clear peaks in the reaction spectra for the absorptive potential strength evaluated in Ref.~\citen{ref:8}.
By scaling the strength of the real part of the potential, we find that we can expect clear peaks in the cross section due to the bound state formation for a strong attractive potential like $|{\rm Re}~V_{\rm opt}^\phi(0,E=m_\phi)|\geq70$ MeV.
We have also considered the sensitivity of the $\phi$-nucleus interaction to the uncertainties of the kaon selfenergies.

As a conclusion, we find that it is difficult to observe clear signals of the $\phi$-mesic nuclei in the standard missing mass observation.
However, the sub-threshold meson production in nucleus has the definite advantage to observe meson decay and absorption at finite density.
Thus, we think that further theoretical investigations are very important to discover the better way to observe $\phi$ inside nucleus.

\label{sec:4}

\section*{Acknowledgements}
This work is partly supported by the contracts FIS2006-03438, FPA2008-00592 from MICINN (Spain), by CSIC and JSPS under the Spain-Japan research Cooperative program, by Grants--in--Aid for scientific research of MonbuKagakusho and Japan Society for the Promotion of Science No. [20540273] (S.H. ), and by the UCM-BSCH contract
GR58/08 910309.
We acknowledge the support of the European Community-Research Infrastructure Integrating Activity "Study of Strongly Interacting Matter" (HadronPhysics2, Grant Agreement n. 227431) under the Seventh Framework Programme of EU.
J.Y. is a Yukawa Fellow and this work is partially supported by Yukawa Memorial Foundation.
D.C. wishes to aknowledge finantial support from the "Juan de la Cierva" Programme of MICINN (Spain).



\end{document}